\documentclass{aa}

\makeatletter
\AddToHook{begindocument/before}{%
  \@ifpackageloaded{lineno}{%
    \nolinenumbers
    \providecommand{\linenumbers}{}%
  }{}%
}
\makeatother

\usepackage{lineno} % ensure it's present under our control
\AtBeginDocument{%
  \nolinenumbers
  \renewcommand{\linenumbers}{}%
}

\usepackage[varg]{txfonts}
\usepackage{comment}
\usepackage{xcolor}
\usepackage{hyperref}        % do NOT load nameref explicitly

\bibpunct{(}{)}{;}{a}{}{,}

\makeatletter
\renewcommand*\aa@pageof{, page \thepage{} of \pageref*{LastPage}}
\makeatother

\newcommand{\astrobrowser}{\textit{AstroBrowser}}

\newcommand{\smallurl}[1]{{\small\url{#1}}}

\begin{document}

\title{Hierarchical Progressive Survey (HiPS) format: Moving from visualisation to scientific analysis}
\titlerunning{HiPS format: From visualisation to science}

\author{Fabrizio Giordano\inst{1,2} \and Yago Ascasibar\inst{1,3} \and Luca Cortese\inst{4} \and Ivan Valtchanov\inst{5} \and Bruno Mer\'{i}n\inst{6}}
\institute{
Departamento de Física Teórica, Universidad Autónoma de Madrid, 28049 Madrid, Spain
\and
Serco for the European Space Agency (ESA), European Space Astronomy Centre (ESAC), Camino Bajo del Castillo s/n, 28692 Villanueva de la Ca\~nada, Madrid, Spain
\and
Centro de Investigación Avanzada en Física Fundamental (CIAFF-UAM), 28049 Madrid, Spain
\and
International Centre for Radio Astronomy Research, The University of Western Australia, 35 Stirling Highway, Crawley, Perth, WA 6009, Australia
\and
Telespazio UK for European Space Agency, ESAC, Camino Bajo del Castillo, 28692 Villanueva de la Ca\~nada, Madrid, Spain
\and
European Space Astronomy Centre (ESAC) Science Data Centre, Camino Bajo del Castillo s/n, 28692 Villanueva de la Ca\~nada, Madrid, Spain
}


\abstract
{
In the current era of multi-wavelength and multi-messenger astronomy, international organisations are actively working on the definition of new standards for the publication of astronomical data, and substantial effort is being devoted to making them available through public archives.
}
{
We present a set of tools that allow user-friendly access and the basic scientific analysis of observations in the Hierarchical Progressive Survey (HiPS) format, and we use them to gauge the quality of representative skymaps at ultraviolet, optical, and infrared wavelengths.
}
{
We apply a fully automatic procedure to derive aperture photometry in ten different bands for the 323 nearby galaxies in the Herschel Reference Sample (HRS), and compare its results with the rigorous analyses involving specialised knowledge and human intervention carried out by the HRS team.
}
{
Our experiment shows that nine of the ten skymaps considered preserve the original quality of the data, and the photometric fluxes returned by our pipeline are consistent with the HRS measurements within a few percent.
In the case of Herschel PACS maps at 100~\textmu m, we uncovered a systematic error that we ascribe to an inconsistent combination of data products with different spatial resolutions.
For the remaining skymaps, the estimated statistical uncertainties provide a realistic indication of the differences with respect to the HRS catalogue.
}
{
In principle, the currently available HiPS skymaps in the Flexible Image Transport System (FITS) format allow one to carry out broadband photometric analyses with an accuracy of the order of a few percent, but some level of human intervention is still required.
In addition to assessing data quality, we also propose a series of recommendations to realise the full potential of the HiPS format for the scientific analysis of large astronomical datasets.
}

\maketitle

\section{Introduction}

Due to the recent advances in the astronomical instrumentation installed on both ground- and space-based observatories, as well as the increased complexity of the data reduction and analysis techniques, it is evermore common for science to be based on observations acquired by large international consortia and exploited not only by the collaboration members but by the whole scientific community.
There is growing awareness of the importance of open science, and public astronomical archives provide an unprecedented wealth of measurements across the entire electromagnetic spectrum.
Their pivotal role in modern astronomy is emphasised in the most recent U.S. Decadal Survey for Astronomy and Astrophysics \citep{NAP26141}, as well as an analysis of the publication record of the European Space Agency (ESA) missions \citep{DeMarchi2024}.
In both cases, it is pointed out that the scientific publications derived from archival data surpass in number those produced by the researchers involved in the original programmes, achieving comparable impact in terms of citations.

However, the exploitation of these complex datasets often requires significant expertise in the specific characteristics of each instrument, its calibration procedures, and data reduction pipelines.
Despite efforts to increase data accessibility, there remains a pressing need for tools that enable scientific analysis without requiring advanced technical skills \citep{Goodman2012}.
This is particularly relevant in the context of multi-wavelength (and multi-messenger) astronomy that involves the analysis and interpretation of extremely heterogeneous datasets.

One important first step in this direction is the definition of a universal standard for storing and distributing astronomical observations.
The Flexible Image Transport System (FITS) format \citep{Wells+81} and the representation of celestial co-ordinates \citep{Calabretta&Greisen2002} in the World Coordinate System \citep[WCS;][]{Greisen&Calabretta02} have been instrumental in this respect.
For large data volumes, the Hierarchical Progressive Survey (HiPS) format \citep{Fernique2015}, developed by the \emph{Centre de Données astronomiques de Strasbourg} (CDS) and the International Virtual Observatory Alliance (IVOA), offers a promising solution to visualise and distribute astronomical observations based on the HTTP protocol.
HiPS organises full-sky data into a multi-resolution tiling scheme based on HEALPix tessellation \citep{Gorski+05}.
It has been adopted by major astronomical institutions such as ESA and ESO, and it is supported by the Aladin software suite \citep{Bonnarel2000}, as well as web applications such as ESASky \citep{Giordano2018}, derived from Aladin Lite \citep{Boch2014}.

Although the HiPS format was conceived with scientific analysis in mind, it has often been used primarily for visualisation based on compressed raster images (e.g. PNG, JPEG), which lack the fidelity and metadata necessary for scientific measurements.
In this approach, researchers seeking quantitative analysis must still retrieve and interpret raw instrument-specific FITS files from archival repositories.
Nevertheless, the number of datasets that are also available in the FITS format is steadily increasing, and here we advocate for the complementary approach of leveraging these skymaps for scientific analysis in addition to the interactive visualisation enabled by raster images.

Combining datasets with varying resolutions, sensitivities, and spectral coverages undoubtedly presents significant challenges, but their integration also enables new scientific insights that transcend individual datasets.
Our goal is not to replace detailed studies requiring deep knowledge about a particular instrument (or set of instruments), but to complement them with accessible, reproducible, multi-wavelength analyses that may be carried out by a much broader segment of the astronomical community.
To that end, we provide a set of tools to retrieve, display, and interact with HiPS datasets from local or remote sources, and to perform basic scientific manipulation (aperture photometry) on a list of targets specified by the user.
In order to gauge the quality of the publicly available skymaps and test the feasibility of recovering accurate broadband photometry by means of a generic pipeline that is not optimised for any particular instrument, we used the Herschel Reference Survey \citep[HRS;][]{Boselli+10} of nearby galaxies as a realistic test case to benchmark our approach.

Section~\ref{sec:observations} summarises the main features of the HRS sample, the flux measurements quoted in the official catalogue, and the HiPS skymaps that are used in the present work.
Section~\ref{sec:analysis} describes the AstroBrowser software suite, its public interface, and its technical details, as well as the data analysis performed on the HiPS skymaps, including the retrieval of the required metadata and the automatic procedure to carry out aperture photometry, background subtraction and error estimates.
A quantitative comparison between our results and the original HRS measurements is presented in Section~\ref{sec:results}, and the main implications in the broader context of the scientific exploitation of HiPS skymaps are briefly discussed in Section~\ref{sec:conclusions}.

\section{Astronomical observations}
\label{sec:observations}

\subsection{Herschel Reference Sample}

The HRS is a volume-limited sample of galaxies with distances between 15 and 25 megaparsecs.
This sample includes 261 late-type galaxies with a 2MASS K-band magnitude of \(K_{S,\ \text{tot}} \leq 12\)~mag, and 62 early-type galaxies with \(K_{S,\ \text{tot}} \leq 8.7\) mag. Detailed information on the initial selection process is available in \citet{Boselli+10}, with updated morphological classifications and distance estimates provided by \citet{Cortese+12}.
Aperture photometry in different bands across the electromagnetic spectrum has been published by \citet{Cortese+12, Cortese+14, Ciesla+12}.
Apertures are generally re-scaled on the optical radius of each target to make sure that the total flux of each source is recovered.
For more details on the procedure, we point the reader to the relevant papers.

The vast majority of galaxies in the HRS sample are clearly spatially resolved from the ultraviolet to the far-infrared regime, and the approach adopted by the HRS team was to treat each object individually, rather than resorting to automatic tools, in order to maximise the accuracy of the multi-wavelength photometric measurements.
Specifically, photometry across the GALEX and SDSS bands was obtained by first fitting surface brightness profiles in elliptical annuli, defined primarily from the SDSS $i$-band images.
The sky background was manually determined in each band by identifying sky-free regions and masking background and foreground sources.
Asymptotic magnitudes were then estimated through a linear fit of the growth curve in each individual band.
Details on the whole procedure are presented in \citet{Cortese+12}.
The approach followed for the Herschel bands was slightly different, primarily due to the additional complexity of the far-infrared observations.
In the SPIRE bands, integrated flux densities for unresolved sources were determined via a timeline-based point-spread function fitting.
For extended sources, aperture photometry was employed, with the size of the aperture rescaled to ensure it encapsulated the entire emission.
This practically implies a morphology-dependent aperture size, with 1.4 times the optical radius (defined as the 25 mag/arcsec$^2$ isophotal radius in the $B$-band) for late-type galaxies, 0.8 times for lenticulars, and 0.3 for ellipticals.
More details are presented in \citet{Ciesla+12}.
For PACS data, photometry was carried out within the same apertures used for SPIRE, except for unresolved SPIRE sources that, being generally resolved in PACS, needed the definition of an elliptical aperture following the same criteria used for SPIRE.
More details can be found in \citet{Cortese+14}.

This sample is ideally suited to test the quality of the available data in HiPS format for several reasons.
First, it makes use of a representative selection of widely used surveys across a broad wavelength range, from the ultraviolet to the far infrared regime.
It is also a volume-limited sample, selected in terms of distance and luminosity, and therefore it includes galaxies of diverse physical properties, such as masses, morphologies, star formation rates, gas and dust content, and environments.
It is composed of nearby, spatially resolved targets, with reliable measurements up to large galactocentric distances in all spectral bands, which facilitates the analysis of very heterogeneous instruments by a common, generic pipeline.
The sample size is small enough to enable the traditional approach based on human expertise, but it perfectly exemplifies the advantages of a more automated procedure;
the level of detail and human intervention used more than a decade ago to produce the HRS measurements makes this sample a fantastic test case with which to assess the performance of an automatic pipeline.

\subsection{HiPS skymaps}

The \textit{Galaxy Evolution Explorer} \citep[GALEX;][]{Martin+05} was a NASA Small Explorer mission designed to study the evolution of star formation in galaxies from the early Universe to the present day.
Using microchannel plate detectors, GALEX captured direct images and low-resolution spectra in the near-ultraviolet (NUV; $\lambda\sim 1344-1786$~\AA) and far-ultraviolet (FUV; $\lambda\sim 1771-2831$~\AA) wavelengths.
Launched in 2003, the mission conducted several surveys, including the All-Sky Imaging Survey (AIS), Medium Imaging Survey (MIS), and Deep Imaging Survey (DIS).
While the FUV camera ceased operations in 2009 due to hardware failure, the NUV instrument continued collecting data until 2012.
The GALEX HiPS skymaps were constructed from the GR6/GR7 data releases \citep{Bianchi+17}, including all the FUV\footnote{\url{https://alasky.cds.unistra.fr/GALEX/GALEXGR6_7_FUV}} and NUV\footnote{\url{https://alasky.cds.unistra.fr/GALEX/GALEXGR6_7_NUV}} images produced by the pipeline described in \citet{Morrissey+07}, which reads the lists of photon positions and arrival times recorded by the detectors, corrects instrumental effects, projects onto sky co-ordinates, and performs flux calibration, background subtraction, and object detection.

The \textit{Sloan Digital Sky Survey} \citep[SDSS;][]{York+00} is a multi-filter imaging and spectroscopic survey conducted with a dedicated 2.5-meter telescope at Apache Point Observatory.
The SDSS imaging camera \citep{Gunn+98} took its first science-quality data the night of 19 September 1998, and it covered a footprint of $14\,055$ square degrees until its last night of science-quality data on 18 November 2009.
\citet{Stoughton+02} described the main aspects of the data analysis pipeline, including basic reduction (e.g. correction for bias, flat field, and data defects such as bad columns or cosmic rays), astrometric and photometric calibration, and object detection.
Starting from the seventh data release \citep[DR7;][]{Abazajian+09}, the \textit{ubercal} algorithm was used to simultaneously solve for the photometric calibration parameters and relative stellar fluxes using overlapping observations within the SDSS.
The HiPS skymaps in the $g$\footnote{\url{https://alasky.cds.unistra.fr/SDSS/DR9/band-g}}, $r$\footnote{\url{https://alasky.cds.unistra.fr/SDSS/DR9/band-r}}, and $i$\footnote{\url{https://alasky.cds.unistra.fr/SDSS/DR9/band-i}} bands used in the present work were derived from the ninth data release \citep[DR9;][]{Ahn+12}, whose main difference with respect to the DR7 imaging pipeline is an improvement in the astrometric calibration. In terms of photometry, the data should be almost identical to the SDSS observations used by \citet{Cortese+12}.

The \textit{Herschel} Space Observatory \citep{Pilbratt+10} was an ESA mission that observed the far-infrared and submillimeter spectrum from 14 May 2009 to 29 April 2013 using three different instruments: PACS \citep[Photodetector Array Camera and Spectrometer;][]{Poglitsch+10}, SPIRE \citep[Spectral and Photometric Imaging Receiver;][]{Griffin+10}, and HIFI \citep[Heterodyne Instrument for the Far-Infrared;][]{deGraauw+10}.
The HRS catalogue reports photometric measurements in the `green' and `red' PACS bands
at 100\footnote{\url{https://skies.esac.esa.int/Herschel/PACS100}} and 160\footnote{\url{https://skies.esac.esa.int/Herschel/PACS160}} \textmu m, as well as the three SPIRE bands (250\footnote{\url{https://skies.esac.esa.int/Herschel/SPIRE250}}, 350\footnote{\url{https://skies.esac.esa.int/Herschel/SPIRE350}}, and 500\footnote{\url{https://skies.esac.esa.int/Herschel/SPIRE500}} \textmu m).
The raw data from both instruments underwent extensive processing through the Herschel Interactive Processing Environment \citep[HIPE;][]{Ott10}, including astrometry, flux calibration, and projection of the scans associated to the different observing modes onto sky co-ordinates.
The HiPS maps were constructed from all the publicly available Level~2.5 data (or Level~2, if no Level~2.5 was available) in the Herschel Science Archive\footnote{\url{http://archives.esac.esa.int/hsa/whsa/}}.
For PACS, these images were produced by the Jscanam algorithm \citep{Gracia-Carpio+15}, and they are suitable for the scientific analysis of both point and extended sources.
For SPIRE, the extended-source calibrated maps were used, whose zero-level is determined by comparison with specially produced \textit{Planck} HFI maps at 545 and 857~GHz.

\section{Data analysis}
\label{sec:analysis}

Nowadays, several platforms provide flexible frameworks
for accessing and analysing astronomical data. The services they offer are often designed to abstract away mission-specific details, enabling users to focus on scientific goals such as multi-wavelength analysis without needing to address differences in data format or instrumentation.
Key examples include metadata services like \textit{SIMBAD}\footnote{\url{https://simbad.cds.unistra.fr/simbad/}}~\citep{Wenger2000} and \textit{NED}\footnote{\url{https://ned.ipac.caltech.edu/}}~\citep{Helou1991}, which allow researchers to query source information by name or co-ordinates. The CDS \textit{Sesame} service\footnote{\url{https://cds.unistra.fr/cgi-bin/Sesame}} integrates both, along with additional resources. For catalogue access, the \textit{VizieR} service\footnote{\url{https://vizier.cds.unistra.fr/}}~\citep{Ochsenbein2000} is an essential reference, hosting a vast collection of published astronomical catalogues widely used by the community.
In terms of scientific visualisation, \textit{Aladin Desktop}~\citep{Bonnarel2000} offers advanced exploration and analysis of HiPS maps, including cut-out services
(see also the \textit{hips2fits}\footnote{\url{https://alasky.cds.unistra.fr/hips-image-services/hips2fits}} web service and associated APIs, also provided by the CDS)
and catalogue overlays. Its web-based counterpart, \textit{Aladin Lite}~\citep{Boch2014}, is extensively used in portals such as ESASky\footnote{\url{https://sky.esa.int/}}, the ESO Science Archive\footnote{\url{https://archive.eso.org/scienceportal/home}}, and JAXA’s JUDO2\footnote{\url{https://darts.isas.jaxa.jp/app/judo2/}}.
NASA/IPAC’s \textit{IRSA Viewer}\footnote{\url{https://irsa.ipac.caltech.edu/irsaviewer/}},
built upon the \textit{Firefly}\footnote{\url{https://github.com/Caltech-IPAC/firefly}} UI library,
provides similar functionalities with an emphasis on infrared data.

Despite the availability of these established tools, our scientific goals required a platform with greater flexibility and extensibility, tailored to the specific requirements of our study. To this end, we developed \astrobrowser, a lightweight yet versatile framework that allows fine control over data handling, seamless integration with analysis workflows, and support for scientific interoperability,
as well as a set of analysis scripts to carry out background subtraction, aperture photometry, and the estimation of measurement errors.

\subsection{AstroBrowser}

\begin{figure}
	\includegraphics[width=.49\textwidth]
	{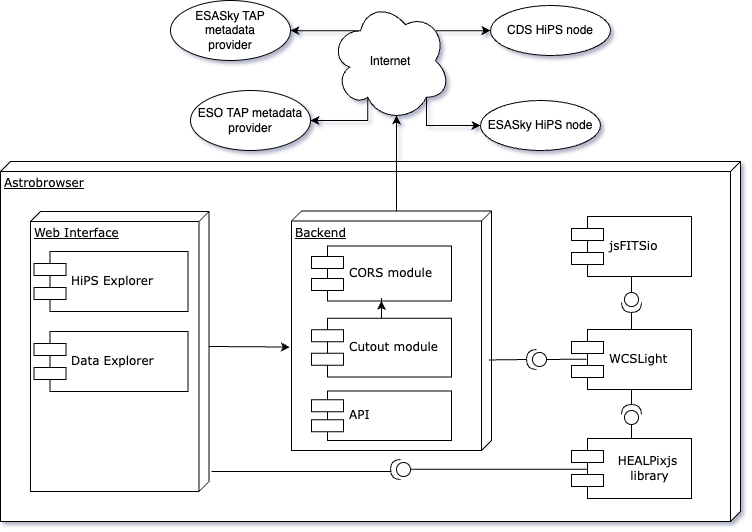}
	\caption{\astrobrowser\ components diagram.}
	\label{fig:components_diagram}
\end{figure}

All the data used in this study were retrieved using \astrobrowser, a modular suite of tools for the acquisition, visualisation, and scientific analysis of astronomical datasets stored in the HiPS format. The system is designed to interface directly with public archives, streamlining access while maintaining consistency with IVOA standards.
The version presented in this work represents the first implementation of a broader, long-term vision aimed at browser-based, supervised collaborative scientific analysis of intermediate-size samples, of the order of hundreds of targets to a few thousand.
The current version implements the core functionality of the framework, and it consists of a web interface\footnote{\url{http://astrobrowser.ft.uam.es}}, a backend, and three independent libraries, as is illustrated in Fig.~\ref{fig:components_diagram} and detailed in the subsections below.
\astrobrowser\ leverages IVOA standards such as the Table Access Protocol (TAP)~\citep{Dowler2011} for metadata access, the Astronomical Data Query Language (ADQL)~\citep{Ortiz2011} for querying external services, and HiPS for the multi-resolution representation of skymaps.
At a lower level, it supports well-established formats and conventions, including FITS for data storage and the WCS~\citep{Greisen2002} for spatial referencing.
The entire source code has been released under the GNU General Public License v3 and is open to community contributions.

The main repository
\footnote{\url{https://github.com/fab77/astrobrowser-vanilla}} hosts both the web interface and the backend.
For the WCSLight, jsFITSio, and HEALPixjs library, both
the source code\footnote{\url{https://github.com/fab77/astrobrowser-vanilla}}
\footnote{\url{https://github.com/fab77/wcslight}}
\footnote{\url{https://github.com/fab77/FITSParser}}
and nodejs packages\footnote{\url{https://www.npmjs.com/package/wcslight}}
\footnote{\url{https://www.npmjs.com/package/jsfitsio}}
\footnote{\url{https://www.npmjs.com/package/healpixjs}}
are provided.

\subsubsection{Web interface}

The web interface of \astrobrowser\ represents the primary user interaction point.
It consists of two different modules:

\begin{figure}
  \includegraphics[width=.49\textwidth]{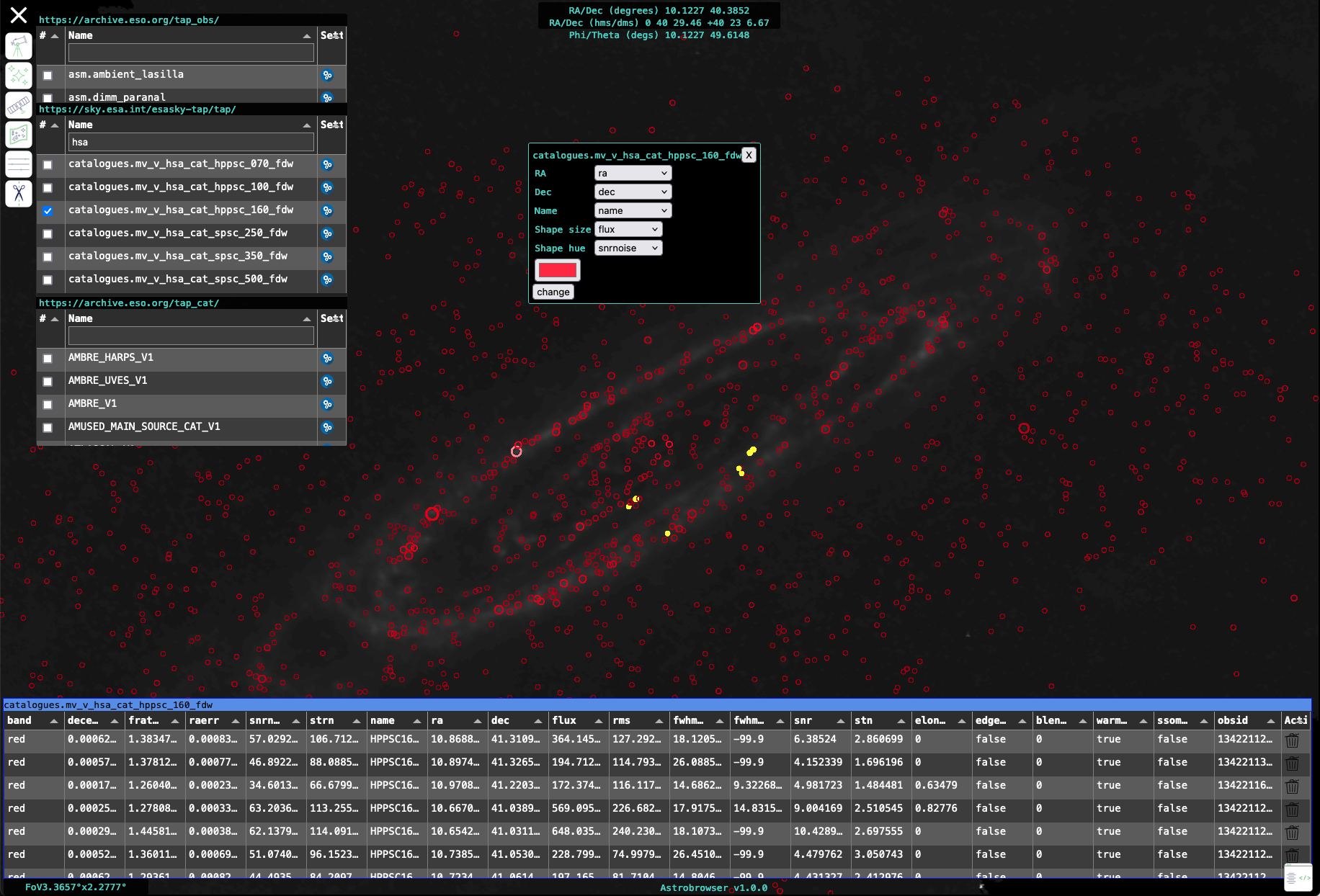}
  \caption{Herschel SPCS 500 catalogue overlay on M31 superimposed on SPIRE 500 HiPS. The shape size is mapped to the \textit{flux} metadata column, and the shape hue is mapped to the \textit{snrnoise} metadata column.}
  \label{fig:astrobrowser_view}
\end{figure}

\textit{HiPS Explorer}, developed in Vanilla JavaScript and WebGL, is the core component that provides an intuitive and visual portal to access all the \astrobrowser\ functionality (see Fig.~\ref{fig:astrobrowser_view}).
It handles data visualisation (e.g. FITS and HiPS maps), and it overlays geometrical metadata representations (e.g. source catalogues and observation footprints) onto an active HiPS map. The list of HiPS maps available in the \textit{HiPS Explorer} is derived from the \textit{Hipslist} file\footnote{\url{https://aladin.cds.unistra.fr/hips/list}} provided by the CDS and ESASky HiPS nodes.
The retrieval of catalogues and observation footprint metadata is performed using ADQL queries on ESASky and ESO TAP service providers.

\begin{figure}
  \includegraphics[width=.49\textwidth]{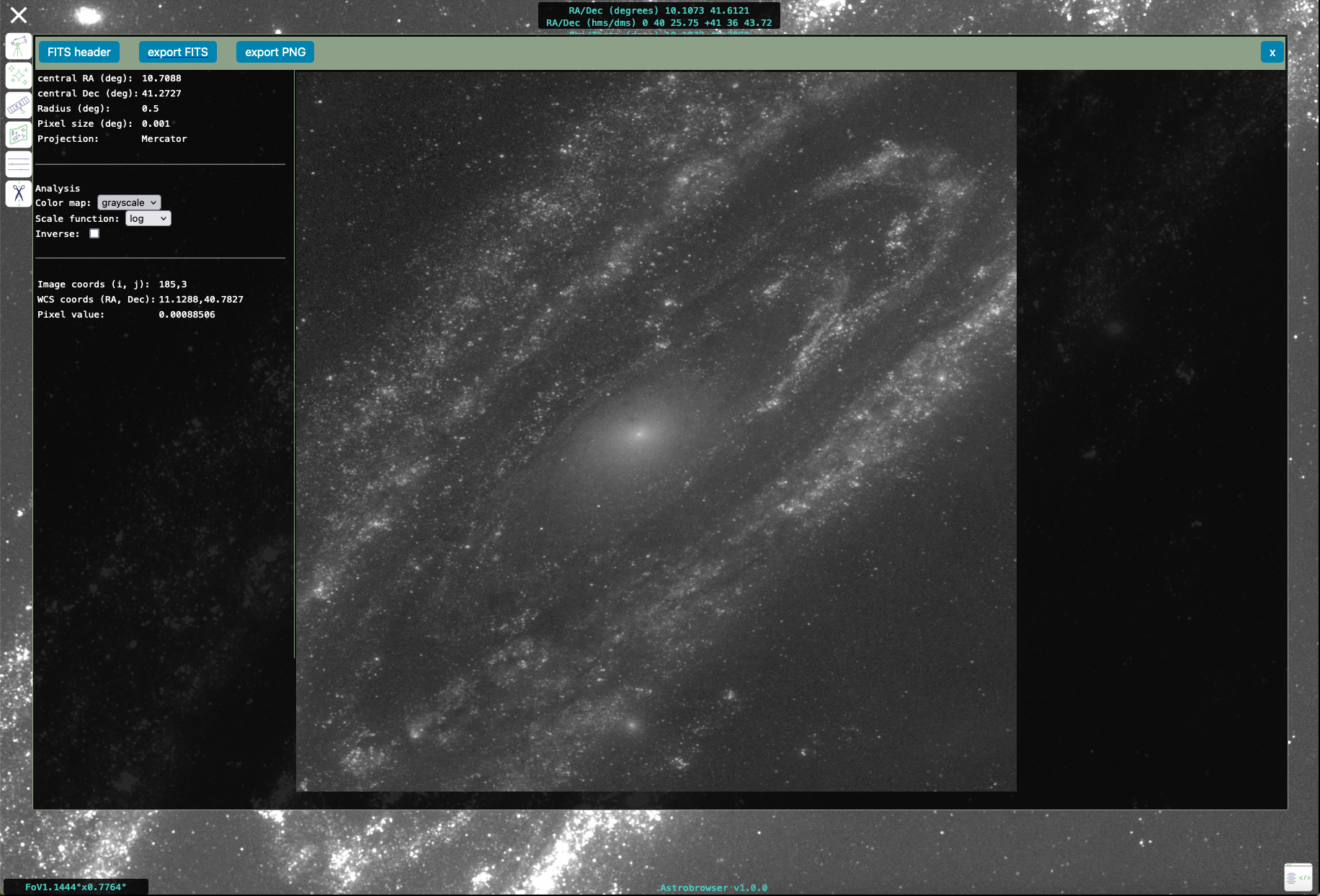}
  \caption{Interactive DataExplorer panel view on GALEX NUV HiPS targeting M31.}
  \label{fig:dataexplorer}
\end{figure}

The \textit{Data Explorer} (Fig~\ref{fig:dataexplorer}) enables a more detailed exploration and manipulation of the data that are currently displayed in the \textit{HiPS Explorer}, and it grants a user-friendly access to the cut-out service exposed in the backend via the API.
Its key features include:
\begin{itemize}
    \item Cut-out extraction: Users can input the central right ascension (RA), declination (Dec), pixel size, and radius of the requested field of view to extract a cut-out from the active HiPS map.
    \item Visual inspection: By simply hovering the mouse, users can explore in real time the physical co-ordinates (i, j), astronomical co-ordinates (RA, Dec), and pixel values of the cut-out image.
    \item Customisation: The colour map and scale function can be adjusted at will to improve visualisation and interactively adapt to the dynamical range of the displayed data.
    \item Download: A postcard image (PNG) and the actual data in FITS format can be downloaded for further analysis or archival purposes.
\end{itemize}

\subsubsection{Backend}

The web interface described above relies on a backend to query external metadata providers, retrieve observational data, and perform various file and data manipulation tasks.
The backend helps one to circumvent cross-origin resource sharing issues when accessing external providers (e.g. HiPS servers and TAP services).
More importantly, it exposes the application programming interface (API) component that facilitates seamless programmatic access to the \astrobrowser\ functionality.
The API consists of a single HTTP entry point that can be called from any program or web service.
Here is an example of how to directly call the cut-out service from a web browser using the API:

\url{http://astrobrowser.ft.uam.es/api/cutout?hipsbaseuri=https://alasky.cds.unistra.fr/GALEX/GALEXGR6_7_NUV/&radeg=10.7&decdeg=41.3&radiusasec=7000&pxsizeasec=15},

where
\begin{itemize}
    \item \texttt{radiusasec} is the desired radius, in arc-seconds;
    \item \texttt{pxsizeasec} the desired pixel size, in arc-seconds;
    \item \texttt{radeg} the central RA, in degrees;
    \item \texttt{decdeg} the central Dec, in degrees;
    \item \texttt{hipsbaseuri} the base URL of a HiPS map, where the properties file is stored.
\end{itemize}

\subsubsection{Stand-alone libraries}

The most common low-level, astronomy-specific operations that are carried out by the different \astrobrowser\ modules have been implemented as three independent libraries that are publicly distributed to help other developers of astronomical software.
They can all be run embedded in a web page or imported as a dependency in any Node.js program.

\textit{jsFITSio} is a TypeScript library that handles I/O operations and data manipulation on standard FITS files.
This library represents the foundational system level in \astrobrowser, enabling the reading and writing of standard FITS files.

\textit{WCSLight} is a TypeScript library designed to manipulate co-ordinate projections according to the FITS standard. It relies on \textit{jsFITSio} to read, write, and download FITS files, and it implements the WCS functions \texttt{pix2world} and \texttt{world2pix}, as well as the low-level cut-out functionality on single FITS files and the entire HiPS.

\textit{HEALPixjs} is a library that partially ports the HEALPix C++ library to Vanilla JavaScript. Since \astrobrowser\ uses NEST tile organisation for HiPS, only the NEST tessellation code has been ported to \textit{HEALPixjs}. This library is used by \textit{WCSLight} for operations on HiPS projections and in the \astrobrowser\ web interface for visualising HiPS maps and supporting the algorithms that allow the selection of footprints overlaid on a HiPS sphere.

\subsection{Analysis pipeline}

For every galaxy in the HRS, we used the \astrobrowser\ web API exposed by the backend to download cut-outs in each band, and then carried out aperture photometry on the retrieved images.
Our analysis pipeline, publicly available as a Python notebook\footnote{\url{https://github.com/paranoya/astrobrowser/blob/main/HRS.ipynb}}, can be described in terms of two separate steps:

\begin{table*}
	\caption{Scientific calibration parameters of each skymap.}
	\label{tab:skymaps_calib}
	\centering
	\begin{tabular}{cccccc}
		\hline\hline
		Band &   HiPS pixel scale     & original pixel scale & PSF FWHM & original units & standardised units\\
		& $s_{\rm pix}$ [arcsec] &    $s_0$ [arcsec]    &    [arcsec]      &                & [$\mu$Jy/arcsec$^2$]\\
		\hline
		GALEX FUV          & $0.805$  & $1.5$   & $4.2$ & counts/s/$s_0^2$ & $47.85$ \\
		GALEX NUV          & $0.805$  & $1.5$   & $5.3$ & counts/s/$s_0^2$ & $14.99$ \\
		SDSS g             & $0.4025$ & $0.396$ & $1.224$ & nanomaggies/$s_0^2$ & $23.20$ \\
		SDSS r             & $0.4025$ & $0.396$ & $1.122$ & nanomaggies/$s_0^2$ & $23.20$ \\
		SDSS i             & $0.4025$ & $0.396$ & $1.071$ & nanomaggies/$s_0^2$ & $23.20$ \\
		Herschel-PACS 100  & $0.805$  & $1.6$   & $6.9$ & Jy/$s_0^2$ & $390703$ \\
		Herschel-PACS 160  & $0.805$  & $3.2$   & $11.9$ & Jy/$s_0^2$ & $97675$ \\
		Herschel-SPIRE 250 & $3.221$  & $6$     & $18.1$ & MJy/sr & $23.50$ \\
		Herschel-SPIRE 350 & $6.440$  & $10$    & $25.2$ & MJy/sr & $23.50$ \\
		Herschel-SPIRE 500 & $12.88$  & $14$    & $36.6$ & MJy/sr & $23.50$ \\
		\hline
	\end{tabular}
\end{table*}

\subsubsection{Calibration}
\label{sec:calibration}

To carry out a meaningful scientific analysis, it is necessary to supplement the information available in the properties file of the different skymaps with some additional data; most notably, the spatial resolution of the original images and the physical units of the intensity values.
In this work, we adopt the parameters quoted on Table~\ref{tab:skymaps_calib}.

The maximum resolution of the HiPS maps can always be inferred from the compulsory \textit{hips\_order} keyword, and it is most often explicitly provided as \textit{hips\_pixel\_scale}.
However, it is common that they oversample the original observations (i.e. they feature smaller pixel size) in order to preserve all the information present in the original data.
This comes, of course, at the cost of losing statistical independence, and one must take into account the original pixel size when computing the number of independent degrees of freedom to estimate the uncertainties, as is explained below.

In the maps considered here, the optional HiPS keyword \textit{s\_pixel\_scale} is provided for most instruments and bands.
For GALEX, though, we had to manually retrieve the information from \citet{Morrissey+07}, while for the $i$ band of the SDSS we decided to assume the same spatial resolution as for the original images in the $g$ and $r$ bands \citep[$0.39564$~arcsec, consistent with the parameters of the SDSS photometric camera;][]{Gunn+98}.
Note that in this case the pixel size of the HiPS maps is slightly larger, and therefore it can be safely assumed that all these measurements are statistically independent.
For the Herschel space observatory, the original pixel sizes reported in the \textit{s\_pixel\_scale} keyword of the HiPS properties files are fully consistent with the official description of the PACS \citep{Poglitsch+10} and SPIRE \citep{Griffin+10} instruments, as well as the online documentation\footnote{\url{https://www.cosmos.esa.int/web/herschel/legacy-documentation}}.
Nevertheless, it is important to highlight that the actual pixel size of the original PACS products at 100~\textmu m depends on the Astronomical Observation Template (AOT): 1.6 arcsec, as is quoted in Table~\ref{tab:skymaps_calib}, for the prime mode, and 3.2 arcsec for parallel mode.

In most cases, the original pixel size, $s_0$, is significantly smaller than the true spatial resolution of the observations, usually characterised in terms of the full width at half maximum (FWHM) of the point spread function (PSF).
Determining the precise shape of the PSF is a complex task, since it depends on both the instrument configuration and the observing conditions of each target and exposure, and there is no straightforward way to define an `average' PSF for a HiPS skymap.
No information in this regard is provided in the properties file, and therefore we simply estimated here the representative values of the typical FWHM of the original observations based on the GALEX~\citep{Morrissey+07}, SDSS\footnote{\url{https://www.sdss4.org/dr16/imaging/other_info/}}, and Hershel \citep{Poglitsch+10, Griffin+10} documentation.
The GALEX PSF is dominated by instrumental effects (most notably, the source position within the detector), whereas in SDSS it is dominated by atmospheric seeing.
The effective beam of the PACS instrument is rather asymmetric, and it depends on the scan speed; since the HiPS skymaps are derived from all the observations available in the archive, the numbers quoted on Table~\ref{tab:skymaps_calib} are merely meant to be representative (i.e. an approximate median) of the precise values reported in \citet{Poglitsch+10} for each configuration.
For SPIRE, we take the mean values of the FWHM provided in \citet{Griffin+10}.

Finding out the precise nature and physical units of the measurements stored in each map also required some manual literature research.
GALEX data contain a \textit{hips\_bunit} keyword in the properties files.
However, such a parameter is not part of the standard recommendation for the HiPS format put forward by the IVOA\footnote{\url{https://www.ivoa.net/documents/HiPS/}}, and the reported values do not coincide with our estimates, quoted in Table~\ref{tab:skymaps_calib}, which assume zero points of $18.82$ (FUV) and $20.08$ (NUV) in the AB system \citep{Oke&Gunn83}, based on \citet{Morrissey+07} and the instrument documentation available online\footnote{\url{https://archive.stsci.edu/missions-and-data/galex}}.
For SDSS and Herschel, flux calibration information was also retrieved from external sources.
SDSS photometry \citep{Padmanabhan+08}, also intended to be on the AB system, is expressed in terms of `nanommaggies', which we approximated as $3.631~\mu$Jy.
According to the Herschel documentation, the standard product generation pipeline provides flux-calibrated PACS images in units of janskys, whereas SPIRE measurements are given in megajanskys per steradian.

Note that only the SPIRE maps contain intensity values.
In all the other cases, the quantity stored in the HiPS maps corresponds to flux per original pixel (\textit{s\_pixel\_scale} $s_0$), regardless of the spatial resolution (\textit{hips\_pixel\_scale} $s_{\rm pix}$) of the HiPS maps.
Therefore, precise knowledge of the original pixel area, $A_0 = s_0^2$, is required in order to convert the map values to intensities per unit solid angle and derive meaningful photometry.
This poses a significant problem for the PACS  100~\textmu m band, where original images with different resolutions (1.6 and 3.2 arcsec per pixel) have been combined without adequate recalibration.

\subsubsection{Aperture photometry}

For each galaxy in the HRS sample, we made use of the \textit{AstroBrowser} backend to download the image cut-outs in every band and compute the average intensity, $\langle I \rangle$, within the elliptical aperture defined by the RA and Dec $(\alpha_0, \delta_0)$ of the galaxy, its semi-major and semi-minor axes $(a, b)$, and the position angle, $\theta$, of the major axis, as provided in the HRS catalogues.
All of our cut-outs were centred around the location of the target, and the field of view was set to eight times the reported semi-major axis, $a$.

To determine the flux received from the galaxy, one must subtract the potential contribution from any large-scale background, be it of astrophysical or instrumental origin.
To do so, we masked an elliptical area three times larger than the region of interest and computed the median and $16-84$ percentiles over the remaining pixels, $I^{\rm out}_{ij}$.
Then, we adopted the median intensity, $I^{\rm out}_{50}$, as indicative of the image background, and our flux estimate is simply
\begin{equation}
F = \pi a b \left( \langle I^{\rm in}_{ij} \rangle - I^{\rm out}_{50} \right),
\end{equation}
where the average $\langle I^{\rm in}_{ij} \rangle$ was performed over the original region of interest (i.e. the ellipse with semi-major and semi-minor axes $a$ and $b$).

No attempt was made to deconvolve the PSF of the different instruments,
since most HRS galaxies are well resolved, and therefore the vast majority of the flux is expected to be contained within the reported elliptical apertures.
The smallest semi-minor axis in the optical/UV \citep{Cortese+12} is 9~arcsec, significantly larger than the PSF of SDSS and GALEX observations.
In the FUV, all semi-minor axes \citep{Ciesla+12, Cortese+14} are greater than 20~arcsec, which also suffices to fully enclose the PACS PSF\footnote{Let us note that the values reported in Table~\ref{tab:skymaps_calib} correspond to the FWHM of the PSF, and therefore they should be compared to the full minor axis $2b$ of the region of interest.}.
SPIRE is the only instrument that is not able to spatially resolve all the galaxies in our sample.
As a matter of fact, 10, 12, and 11 objects were treated as point-like sources by \citet{Ciesla+12} in the 250, 350, and 500~\textmu m bands, respectively.
Thus, only a small fraction (about 5 percent) of our targets would require a detailed modelling of the PSF.

To estimate the measurement uncertainty,
\begin{equation}
\Delta F = N \sqrt{ (\Delta I)^2 + (\Delta B)^2 },
\end{equation}
we added in quadrature the error term $(\Delta I)^2$ associated with the mean intensity within the region of interest and the estimated uncertainty of the background $(\Delta B)^2$, including both systematic and statistical errors, and then multiplied by the total number, $N$, of pixels in the region of interest.

For the mean target intensity, we estimated
\begin{equation}
(\Delta I)^2 = \frac{ (I^{\rm in}_{50} - I^{\rm in}_{16})^2 } { n - 1 }
\end{equation}
from the median and the 16 percentile of the intensity within the target elliptical aperture, where $n = N \min(\frac{ s_{\rm pix}^2 }{ s_0^2 }, 1) \simeq \frac{ \pi a b }{ \max(s_{\rm pix}^2, s_0^2) }$ denotes the number of statistically independent degrees of freedom, based on the number, $N$, of pixels, their area, $A_{\rm pix} = s_{\rm pix}^2$, and the resolution, $A_0=s_0^2$, of the original images (see Table~\ref{tab:skymaps_calib}).

\begin{figure*}
	\centering
	\includegraphics[width=.42\textwidth]{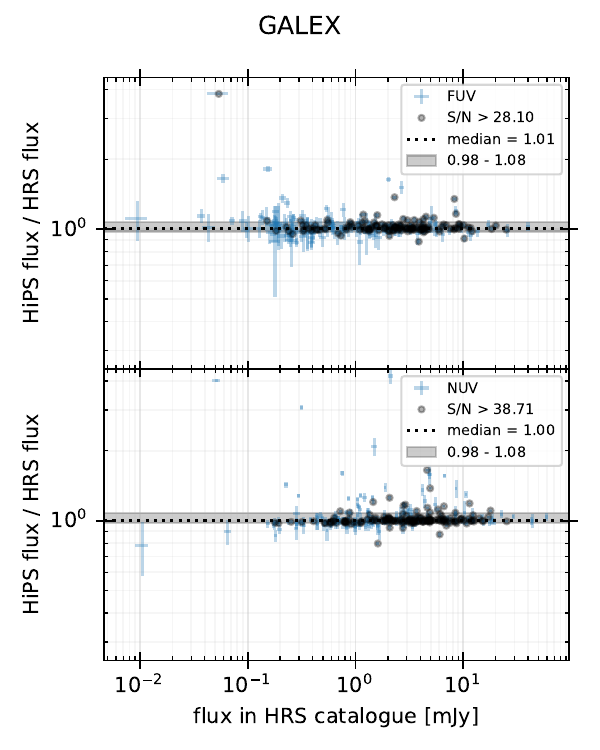}
	\includegraphics[width=.42\textwidth]{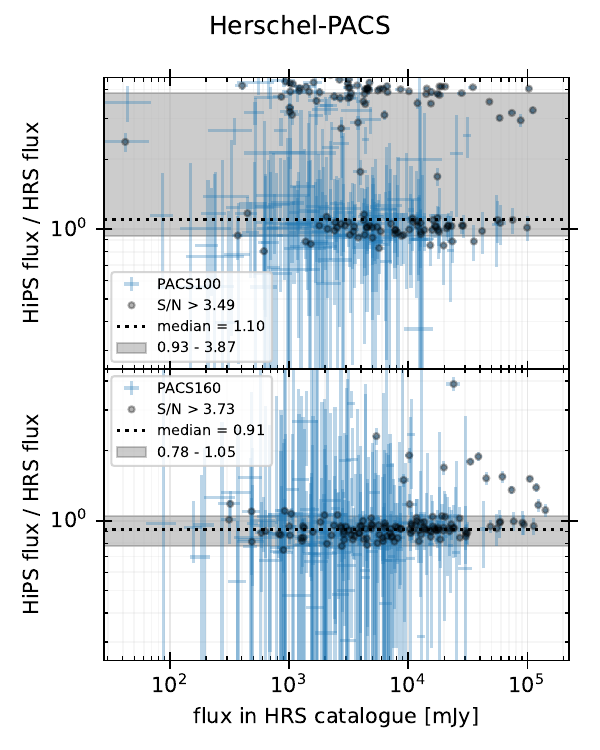}\\
	\includegraphics[width=.42\textwidth]{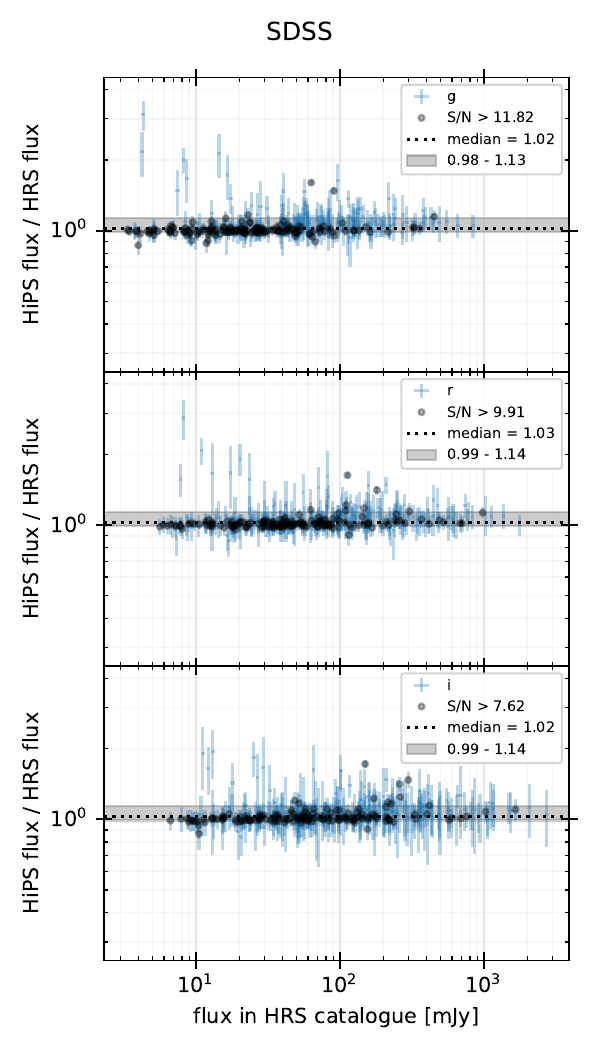}
	\includegraphics[width=.42\textwidth]{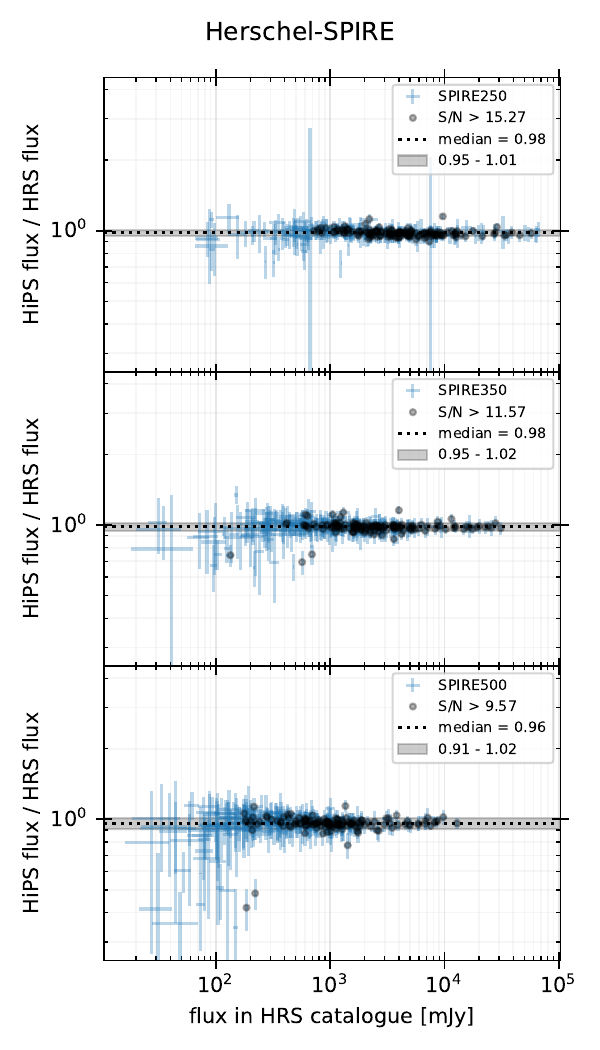}
	\caption{
		Comparison between the flux measured from the HiPS skymaps and the values reported in the HRS catalogues for GALEX \citep[top left]{Cortese+12}, SDSS \citep[bottom left]{Cortese+12}, and Herschel \citep[top and bottom right for the PACS and SPIRE instruments, respectively]{Ciesla+12, Cortese+14}.
			Error bars in the vertical axis represent the uncertainties returned by our pipeline (they do not include HRS errors); measurements above the median $S/N$ are highlighted in black.
			Horizontal dotted lines and grey bands denote the median and $16-84$ percentiles of the flux ratio (see Figure~\ref{fig:flux_comparison} for the full probability distribution).
	}
	\label{fig:fluxes}
\end{figure*}

In order to account for possible systematic uncertainties associated with our background estimate, $I^{\rm out}_{50}$, we made an approximate model of large-scale diffuse emission in the following way. First, we excluded, in addition to the enlarged elliptical mask, all pixels $I^{\rm out}_{ij} > I^{\rm out}_{50} + 1.5\,(I^{\rm out}_{84} - I^{\rm out}_{16})$ above a three-sigma threshold in order to discard nearby bright sources.
We thus obtained a binary mask, $m_{ij}$, and a masked image, $b_{ij} = m_{ij} I_{ij}$, over the whole field of view that isolates the background-dominated area.
In order to interpolate the contribution of the diffuse background within the region of interest, we carried out a weighted Gaussian smoothing with radius $R = \frac{a + b}{4}$:
\begin{equation}
    B_{ij} = \frac{ G_R(b_{ij}) }{ G_R(m_{ij}) },
\end{equation}
where $G_R$ denotes the Gaussian smoothing operation with kernel radius $R$.
Based on this spatially varying model,
\begin{equation}
    (\Delta B)^2 = \frac{ (I^{out}_{50} - I^{out}_{16})^2 } {n} + \left\langle ( B^{in}_{ij} - I^{\rm out}_{50} )^2 \right\rangle
\end{equation}
attempts to account for the statistical uncertainty in the estimate of the mean background intensity (first term) as well as the systematic errors arising from fluctuations of the background model throughout the region of interest (second term).

\section{Results}
\label{sec:results}

As can be seen in Figure~\ref{fig:fluxes}, there is overall good agreement between the values obtained by the procedure above, applied to the publicly available HiPS maps, and the official values provided by the HRS catalogue, carefully supervised by human experts.
The most notable exception is the PACS 100~$\mu$m band, where a clearly visible systematic error is affecting a fraction of the data.
As is noted in Section~\ref{sec:calibration}, we believe that the source of the observed offset may be related to the ingestion of original images with different pixel size, but further investigation is required in order to fully clarify this issue.
In all other cases, the values recovered automatically from the HiPS maps are fairly close, for the vast majority of galaxies, to the fluxes derived from a thorough analysis by experts.
We have verified by visual inspection that most catastrophic failures (discrepancies larger than a factor $1.5$) are due to nearby sources (e.g. foreground stars), extended galactic emission, or complex backgrounds near the edge of the region of interest.
Background estimation is particularly relevant at a low signal-to-noise ratio (S/N), usually associated with faint objects, but we also noticed a few instrumental artefacts in some cases, especially in the PACS160 band, which may affect bright targets as well.
In the SPIRE maps at 500~\textmu m, one can see that significant flux (sometimes more than $50\%$) may be missed due to the extended wings of the PSF for a few objects with $b \lesssim 30$~arcsec.
For the other two SPIRE bands at 250 and 350~\textmu m, the effect is only seen for point-like sources ($b = 20$~arcsec in the catalogue), and it typically amounts to $15-20\%$.

\begin{figure}
	\centering
	\includegraphics[width=.49\textwidth]{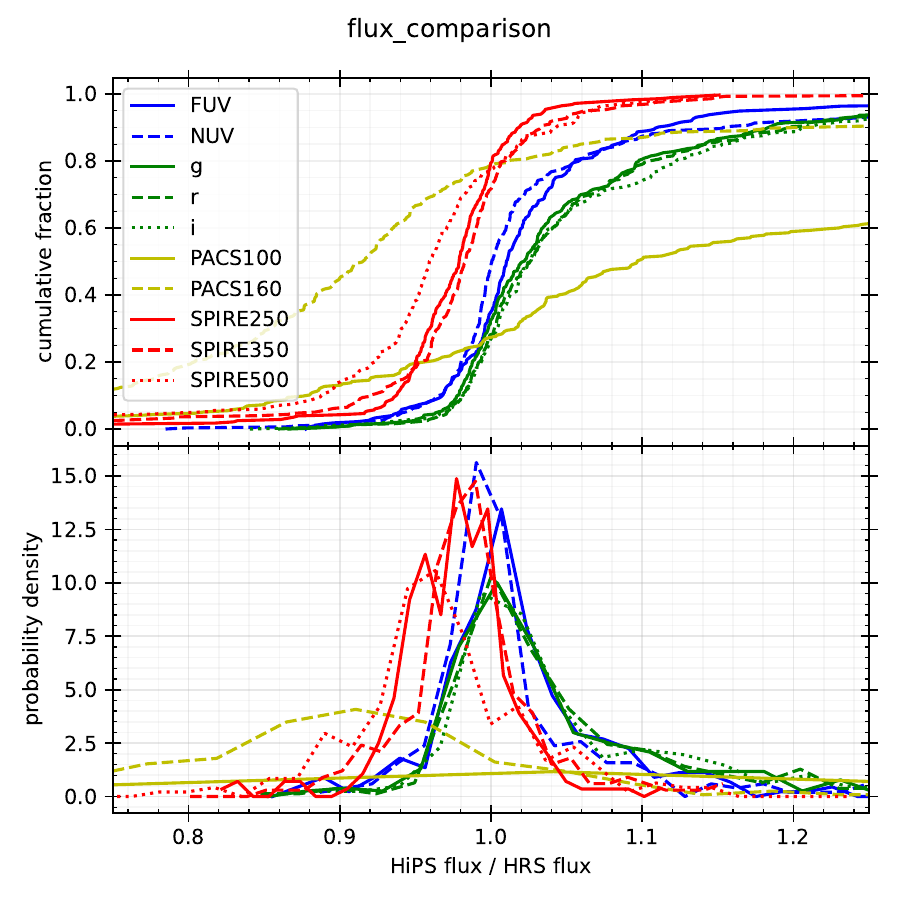}
	\caption{Probability distribution of the ratio between the flux measured from the HiPS skymaps and the values reported in the HRS catalogues. Cumulative fractions and differential probability densities are shown in the top and bottom panels, respectively.}
	\label{fig:flux_comparison}
\end{figure}

A more quantitative assessment of the photometric accuracy of the public HiPS skymaps is provided by the probability distribution of the ratio between the fluxes obtained by our pipeline and the official ones.
Cumulative fractions and differential probability densities for every band are shown in Figure~\ref{fig:flux_comparison}.

The median and $16-84$ percentiles, shown by the dotted lines and grey bands in Figure~\ref{fig:fluxes}, are quite robust to the presence of outliers discussed above, which represent a small fraction of the sample and therefore only affect the extremes of the distribution.
From our analysis, one may conclude that, barring the systematic effects associated to the PACS instrument, the automatic photometry based on the HiPS maps is consistent with the expert analysis within a relative accuracy of a few percent.

With the aforementioned exception, the probability distribution is fairly symmetric for all datasets, and there are very few outliers where the discrepancies are larger than 20 percent.
The median ratio, $p_{50}$, is always close to one, and we do not find evidence of strong systematic biases.
The largest offsets are found for both Herschel PACS bands, but even in this case there is no obvious trend with respect to object brightness.
The width of the distribution is consistent with the typical S/N of the observations (black dots in Figure~\ref{fig:fluxes}). It is also quite stable as a function of observed flux, especially for high-S/N data.
However, our measurements tend to be larger than the HRS values in the optical and the UV, and lower in the FUV.
These trends are responsible for the median offsets and asymmetric tails in the probability distributions shown in Figure~\ref{fig:flux_comparison}.
They are much more evident at low S/N, but they seem to affect the whole sample, and we conjecture that they arise from the subtle differences between our analysis procedure and the prescriptions followed by the HRS team.

\begin{figure}
    \centering
    \includegraphics[width=.49\textwidth]{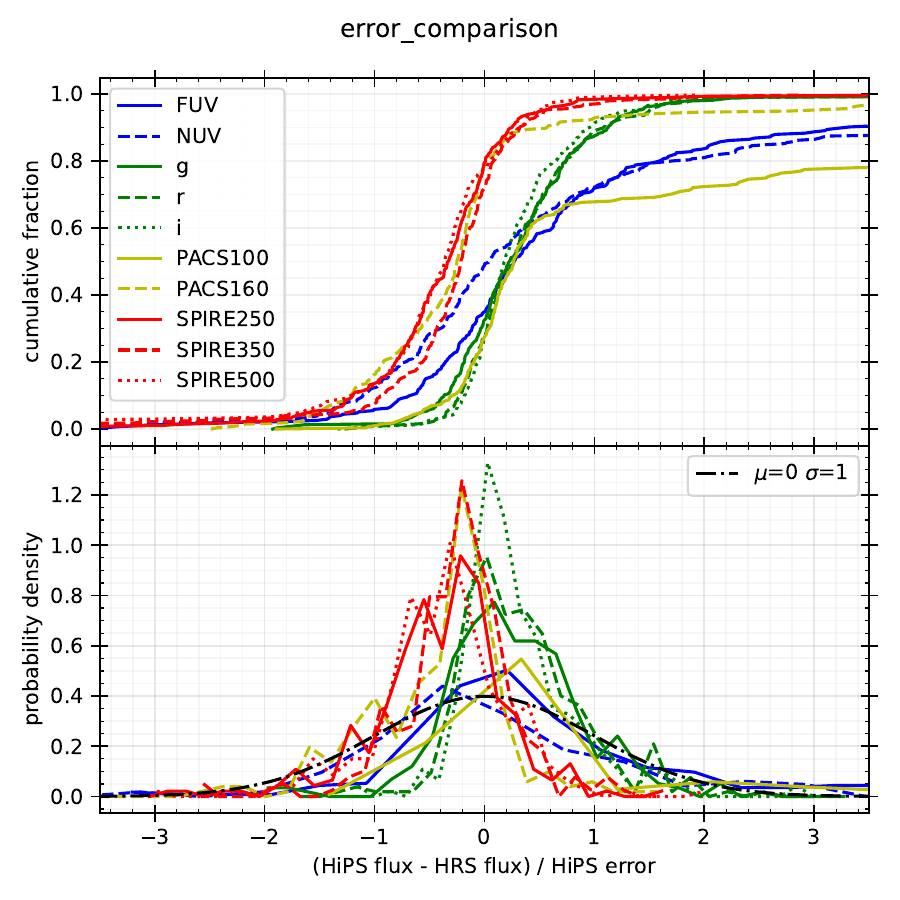}
    \caption{Cumulative fraction (top) and probability density (bottom) of the normalised residual $\chi$ defined in equation~\eqref{eq:chi}. The dash-dotted black line in the bottom panel corresponds to a standard normal distribution.}
    \label{fig:error_comparison}
\end{figure}

\begin{table}
	\caption{Percentiles of the probability distribution of the normalised residual $\chi$, defined in equation~\eqref{eq:chi}.}
	\label{tab:errors}
	\centering
	\begin{tabular}{crrrrr}
	\hline\hline
	Band & $p_{1}$ & $p_{16}$ & $p_{50}$ & $p_{84}$ & $p_{99}$ \\ \hline
	FUV & -3.16 & -0.47 & 0.27 & 2.00 & 10.68\\
	NUV & -3.99 & -0.87 & 0.04 & 2.32 & 11.84\\
	g & -1.71 & -0.23 & 0.22 & 0.85 & 2.22\\
	r & -0.89 & -0.14 & 0.26 & 0.84 & 2.11\\
	i & -0.83 & -0.10 & 0.18 & 0.76 & 1.81\\
	PACS100 & -0.99 & -0.16 & 0.27 & 4.91 & 12.05\\
	PACS160 & -2.23 & -1.01 & -0.25 & 0.13 & 6.17\\
	SPIRE250 & -3.09 & -0.90 & -0.34 & 0.07 & 1.28\\
	SPIRE350 & -4.48 & -0.70 & -0.21 & 0.15 & 1.74\\
	SPIRE500 & -12.50 & -0.90 & -0.36 & 0.13 & 0.83\\
	\hline
	\end{tabular}
\end{table}

Finally, we evaluated the accuracy of the uncertainties returned by our algorithm by studying the probability distribution of the normalised residual
\begin{equation}
\chi = \frac{ F_{\rm HiPS} - F_{\rm HRS} }{ \sigma_{\rm HiPS} },
\label{eq:chi}
\end{equation}
where $F_{\rm HRS}$ denote the catalogued fluxes, whereas $F_{\rm HiPS}$ and $\sigma_{\rm HiPS}$ are the fluxes and errors returned by our pipeline, respectively.
Ideally, the normalised residual, $\chi$, should follow a random normal distribution as closely as possible.
The results shown in Figure~\ref{fig:error_comparison} and the percentile values reported in Table~\ref{tab:errors} suggest that this is arguably not a very good approximation, but the uncertainties derived from the HiPS maps are representative of the true errors.

\section{Discussion and conclusions}
\label{sec:conclusions}

The analysis presented in this work demonstrates the potential of publicly available skymaps in a HiPS format to serve not only as a visualisation tool to retrieve the original data but also as a valuable resource for scientific analysis.
In order to quantify this statement, we compared the fluxes measured by a simple automatic pipeline from the GALEX, SDSS, and Herschel HiPS skymaps with those reported in the HRS catalogue, derived from a careful analysis of a relatively small sample of 323 galaxies carried out by human experts.
Our results reveal an overall good agreement, with discrepancies generally within a few percent. 

On the other hand, they also highlight the need for a few improvements in order to guarantee the reliability and optimise the scientific exploitation of these datasets:

To be fully leveraged in scientific applications, it is essential that the metadata of the HiPS skymaps include a better description of the measurements and their physical units.
While there is no obvious prescription to do so in the properties file, we note that the HiPS standard defined in the official IVOA recommendation allows for the provision of a metadata file in FITS format where this information can be clearly stated in the header.
We encourage map creators to actively make use of these mechanism, preferably through the universal content descriptors and the \textit{BUNIT} keyword.
Given the hierarchical (multi-resolution) nature of the HiPS format, the potential variety of use cases, and the fact the target users are the broad scientific community, we strongly recommend reporting the proper surface brightness (flux per unit solid angle) rather than the flux `per beam', `per [original] pixel', or any other unit that requires specific knowledge about the instrument.

There are several other aspects that are relevant in many scientific applications.
Regarding the energy response of the instrument, we recommend storing the filter ID of the SVO filter service in the \textit{FILTER} keyword.
To provide some information about the PSF, one could follow the convention, widely adopted in radio astronomy, of using the keywords \textit{BMIN}, \textit{BMAJ}, and \textit{BPA} to specify its shape (FWHM) and orientation.
Although our results show that an approximate estimate of the measurement uncertainties can be derived from the data themselves, a more rigorous statistical analysis could be carried out if additional skymaps describing the uncertainties and their covariance were available.

All these aspects impose an additional burden on the map creators in terms of both work and responsibility, but we sincerely think that they would greatly increase the scientific and legacy value of the HiPS skymaps.
In order to facilitate the calibration process, we suggest the definition of reference datasets to benchmark the quality of both the skymaps and their metadata, as we have done here based on the HRS.
Such a system would provide users with an objective measure to help them evaluate whether the publicly available skymaps are suitable for a particular science case.

To that end, the details of map creation, its associated uncertainties, usage instructions, and recommendations, a discussion of potential caveats and limitations of the data, and the results of quality assessment against a recognised reference sample should all be well documented.
In our opinion, formal validation based on peer review would be extremely helpful, and we strongly advocate for the publication of manuscripts dedicated entirely to discussing these issues, without any further scientific content.

In conclusion, we claim that the HiPS standard does not only provide an excellent means to optimise the visualisation of astronomical observations, but a robust, scalable framework to distribute science-grade data.
With the exception of the PACS maps at 100~$\mu$m, all the other maps considered in the present work are perfectly suitable for photometric studies that do not critically depend on an accuracy better than $\sim 10\%$.
While no automated procedure will ever be able to achieve the same quality standards as a careful analysis by seasoned human experts, HiPS skymaps represent an invaluable resource for the broad astronomical community, especially in the context of multi-wavelength and multi-messenger studies that must combine measurements carried out by a wide variety of instruments.

\begin{acknowledgements}

We thank the anonymous referee for a constructive report that has helped us improve the presentation of our motivation and the discussion of the results.

Y. Ascasibar received financial support from the Spanish State Research Agency (AEI/10.13039/501100011033) during the inital stages of this work through grant PID2019-107408GB-C42.
A significant fraction of the project was carried out during a research visit to the International Centre for Radio Astronomy Research -- University of Western Australia (ICRAR-UWA) funded from the mobility programme \emph{Ayudas de recualificación del profesorado universitario funionario o contratado} (CA2/RSUE/2021-00817) of the \emph{Universidad Autónoma de Madrid} (UAM), within the framework of the \emph{Plan de Recuperación, Transformación y Resiliencia} of the \emph{Ministerio de Universidades} (Spain).

This work made use of Astropy:\footnote{\url{http://www.astropy.org}} a community-developed core Python package and an ecosystem of tools and resources for astronomy \citep{astropy:2013, astropy:2018, astropy:2022}, as well as the Numpy \citep{harris2020array} and Scipy \citep{2020SciPy} Python libraries.

Funding for SDSS-III has been provided by the Alfred P. Sloan Foundation, the Participating Institutions, the National Science Foundation, and the U.S. Department of Energy Office of Science. SDSS-III\footnote{\url{http://www.sdss3.org}} is managed by the Astrophysical Research Consortium for the Participating Institutions of the SDSS-III Collaboration including the University of Arizona, the Brazilian Participation Group, Brookhaven National Laboratory, Carnegie Mellon University, University of Florida, the French Participation Group, the German Participation Group, Harvard University, the Instituto de Astrofisica de Canarias, the Michigan State/Notre Dame/JINA Participation Group, Johns Hopkins University, Lawrence Berkeley National Laboratory, Max Planck Institute for Astrophysics, Max Planck Institute for Extraterrestrial Physics, New Mexico State University, New York University, Ohio State University, Pennsylvania State University, University of Portsmouth, Princeton University, the Spanish Participation Group, University of Tokyo, University of Utah, Vanderbilt University, University of Virginia, University of Washington, and Yale University.
\end{acknowledgements}

\bibliographystyle{aa}
\bibliography{biblio.bib}

\end{document}